\documentclass[useAMS,usenatbib]{mn2e}
\usepackage{url}
\usepackage{graphicx}
\title[The nature of the cataclysmic variable PT Per]{The nature of the cataclysmic variable PT Per}
\author[M.G.Watson et al.]{M.G.Watson$^1$\thanks{e-mail: mgw@le.ac.uk}, A.Bruce$^2$, C.MacLeod$^2$, J.P.Osborne$^1$ \& A.D.Schwope$^3$\\
$^1$ University of Leicester, Leicester, United Kingdom\\
$^2$ Institute for Astronomy, University of Edinburgh, United Kingdom\\
$^3$ Leibniz-Institut f{\" u}r Astrophysik Potsdam (AIP), Potsdam, Germany}

\begin{document}

\date{}

\pagerange{\pageref{firstpage}--\pageref{lastpage}} \pubyear{2002}

\maketitle

\label{firstpage}

\begin{abstract}
We present a study of the cataclysmic variable star PT Per based on archival XMM-Newton X-ray data and new optical spectroscopy from the WHT with ISIS. The X-ray data show deep minima which recur at a period of 82 minutes and a hard, unabsorbed X-ray spectrum. The optical spectra of PT Per show a relatively featureless blue continuum. From an analysis of the X-ray and optical data we conclude that PT Per is likely to be a magnetic cataclysmic variable of the polar class in which the minima correspond to those phase intervals when the accretion column rotates out of the field of view of the observer. We suggest that the optical spectrum, obtained around 4 years after the X-ray coverage, is dominated by the white dwarf in the system, implying that PT Per was in a low accretion state at the time of the observations. An analysis of the likely system parameters for PT Per suggests a distance of $\approx90$ pc and a very low-mass secondary,  consistent with the idea that PT Per is a ``period-bounce" binary. Matching the observed absorption features in the optical spectrum with the expected Zeeman components constrains the white dwarf polar field to be $B_p\approx 25-27$ MGauss.

\end{abstract}


\section{Introduction}
The XMM-Newton serendipitious survey catalogues \citep[2XMM; 3XMM, ][]{Watson09,ros16} and their associated data (spectra and time series for individual X-ray sources) provide a rich resource for exploring both Galactic and extragalactic source populations down to relatively faint X-ray fluxes ($\sim10^{-14}\mathrm{\ erg\ cm^{-2}\ s^{-1}}$). The most recent catalogue (3XMM-DR5) covers around 2\% of the sky and contains spectra and time series for 24\% of the $\approx 400000$ sources (Rosen et al., 2015). A significant fraction of the brighter X-ray sources ($f_{\mathrm X}>10^{-12}\mathrm{\ erg\ cm^{-2}\ s^{-1}}$) in the 2XMM and more recently the 3XMM catalogues have been visually examined to look for both data processing problems and also to find astrophysically interesting objects. In this paper we present the discovery of periodic eclipse-like features in the X-ray light curve of the cataclysmic variable (CV) PT Per which was in the field of view of an XMM-Newton observation made in 2011 and discuss the nature of this system.

PT Per is a relatively obscure and little studied cataclysmic variable (CV). It is listed in the \citet{downes01} catalogue  as a U Gem (dwarf nova) system with magnitude in the range $\sim$15 - 18.5$^{\mathrm{m}}$. The only publication to present observations of PT Per is \citet{bruch92} which contains an optical spectrum obtained at Calar Alto in 1988. The relatively poor spectrum shows a red continuum with Balmer lines in absorption. The authors note that the spectrum does not resemble that of a CV and comment that it might be a reddened B-star. The possibility also exists that the wrong star was observed as noted by \citet{downes01}.

\begin{figure*}
\includegraphics[width=18cm]{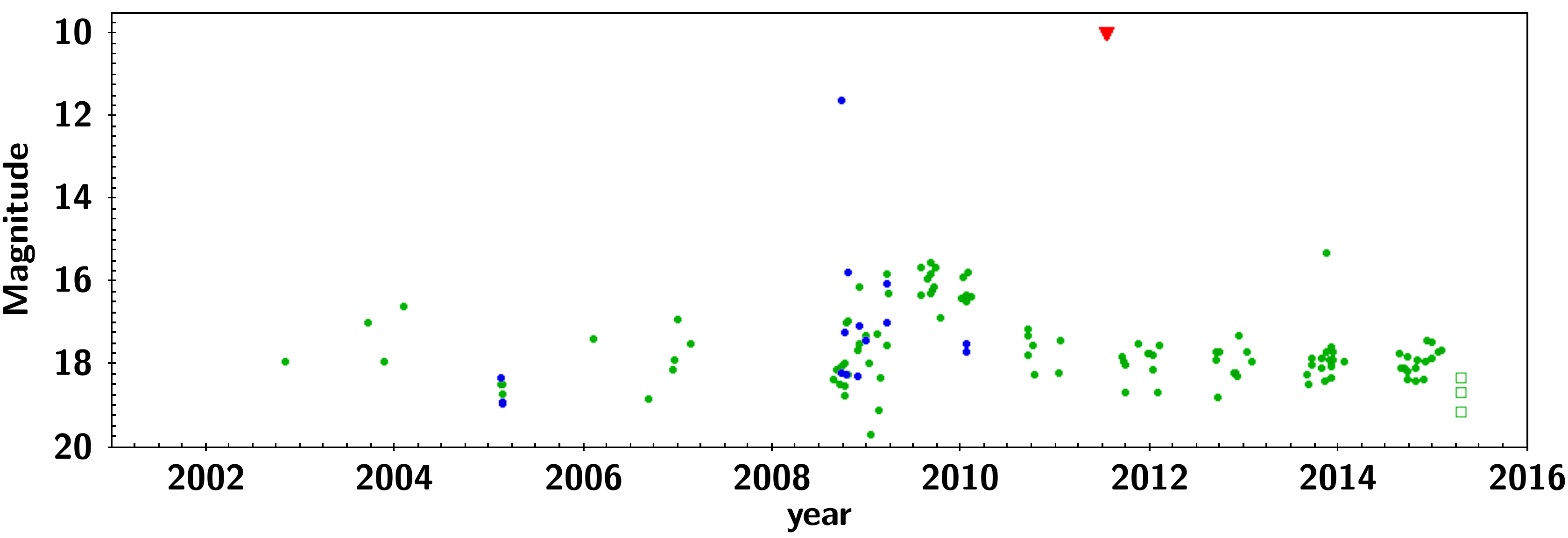}
\caption{AAVSO optical light curve of PT Per. Only V-band (green symbols)and B-band (blue symbols) measurements are shown. The red triangle indicates the date of the XMM-Newton observation. The three open green squares indicate the V-band magnitude of PT Per derived from the WHT spectroscopy (Table 1).}
\end{figure*}

Fig.1 shows the long-term optical light curve of PT Per extracted from the AAVSO observations\footnote{\url{http://www.aavso.org}}. Only V-band and B-band data are shown, thus excluding a number of other measurements in other bands or with no quoted band. Upper limits are also excluded from the plot. These excluded measurements do not provide a significantly different picture of the long-term variability. As can be seen PT Per displays a possible outburst reaching B$\approx 12^{\mathrm m}$ in late 2008 with an amplitude $\approx6^{\mathrm{m}}$. The subsequent brightness appears to be elevated for an interval of over a year and a second, smaller amplitude possible outburst is evident in late 2013. {The nature of these two outbursts is unclear however. The first potential outburst was preceded by an upper limit  (not shown) obtained $\sim 2$ h earlier (implying a very fast rise time) and a measurement consistent with quiescence one day later, whilst the second is bracketed by measurements made 3 days earlier and later both consistent with quiescence.  The maximum duration of the outbursts, assuming they are not erroneous measurements, is thus one day and three days respectively. Alternatively these outbursts might correspond to short duration flares rather than typical CV outbursts, such as has been seen for example in the polar UZ For \citep{pandel02}.

Outside the outbursts the system shows apparently erratic variability between V$\approx19^{\mathrm m}$ and V$\approx 16^{\mathrm m}$.  The size of outburst shown in the AAVSO light curve  is not really consistent with this object being a dwarf nova. \citet[][his Fig.3.8]{warner03} shows that dwarf nova outbursts with this amplitude recur on 100-1000 day timescales. Recurrent outbursts in PT Per are not evident in the long-term light curve. The classification of PT Per as a dwarf nova does not seem to be supported by the AAVSO data. {Note that PT Per has not been covered by the SuperWasp project.\footnote{\url{http://exoplanetarchive.ipac.caltech.edu/docs/SuperWASPMission.html}}

All magnitudes in this paper are on the Vega system. Times quoted are {\bf not} converted to the solar system barycentre.

\section{Observations}
\subsection{X-ray observations}
PT Per was detected serendipitously in an XMM-Newton observation of HD16691 (OBSID 0671100101) made on 2011 July 21 with a total exposure time of 17667 seconds. The 3XMM catalogue source  3XMM J024251.2+564131, at RA(J2000)=02$^h$ 42$^m$ 51.21$^s$, Dec.(J2000)=56$^\circ$ 41' 31.6" (1$\sigma$ position error=0.65") lies close to a chip gap in the EPIC pn CCD camera and outside the field of view of the EPIC MOS1 camera. Nevertheless the proximity to the chip gap is fully accounted for in the analysis and does not affect the observations presented in this paper. This part of the field is also not covered by the XMM-Newton Optical Monitor, precluding any simultaneous optical/UV measurements. The source lies at an off-axis angle of $\sim 13$ arcmin. and is detected with $\approx 2500$ total counts in the EPIC pn and EPIC MOS2 cameras combined, corresponding to a flux of $3.49\times 10^{-12}\mathrm{\ erg\ cm^{-2}\ s^{-1}}$ in the 0.2-12 keV band (assuming a standard power-law spectrum, see Watson et al., 2009).  The position of the 3XMM source is 0.3 arcsec. from the optical coordinates of PT Per, making the identification secure. Fig.2 shows the X-ray position of 3XMM J024251.2+564131 overlaid on the DSS blue image of the region around PT Per.

PT Per was also detected by ROSAT in its all-sky survey \citep{voges99, voges00}. The ROSAT All Sky Survey (RASS) count rate is  0.04 PSPC counts s$^{-1}$, corresponding to an unabsorbed X-ray flux of  $\approx 7.2\times 10^{-13}\mathrm{\ erg\ cm^{-2}\ s^{-1}}$ in the 0.1-2.4 keV band for an assumed  power-law spectrum with $\Gamma\approx 1.7$ and $N_{\mathrm{H}}\approx 3\times10^{20}\ \mathrm{cm^{-2}}$. This is around a factor five fainter than observed by XMM-Newton. The observations  (1990 Aug 7 to Aug 9) occurred 18 years before the optical outburst visible in Fig.1 and 21 years before the XMM-Newton observation. The ROSAT hardness ratios for the detection of PT Per are $\mathrm{HR1}=0.68\pm0.11$ and $\mathrm{HR2}=0.57\pm0.2$\footnote{ROSAT hardness ratios are defined as (B-A)/(A+B) where for HR1, A=0.1-0.4 keV, B=0.5-2 keV whilst for HR2, A=0.5-0.9 keV and B=0.9-2 keV.}
indicating a hard spectrum absorbed X-ray source. These hardness values are only consistent with very flat power-law spectra, $\Gamma\approx 0$ and $N_{\mathrm{H}}>10^{21}\ \mathrm{cm^{-2}}$ or high temperatures ($kT>15$ keV) for a thermal spectrum. For these approximate spectral parameters, the implied unabsorbed ROSAT flux in the 0.1-2.4 keV band is $\ge 9.4\times 10^{-13}\mathrm{\ erg\ cm^{-2}\ s^{-1}}$. {No X-ray coverage of PT Per exists within the Swift archive.\footnote{eg. \url{http://www.swift.ac.uk}}

\begin{figure}
\includegraphics[width=6.5cm]{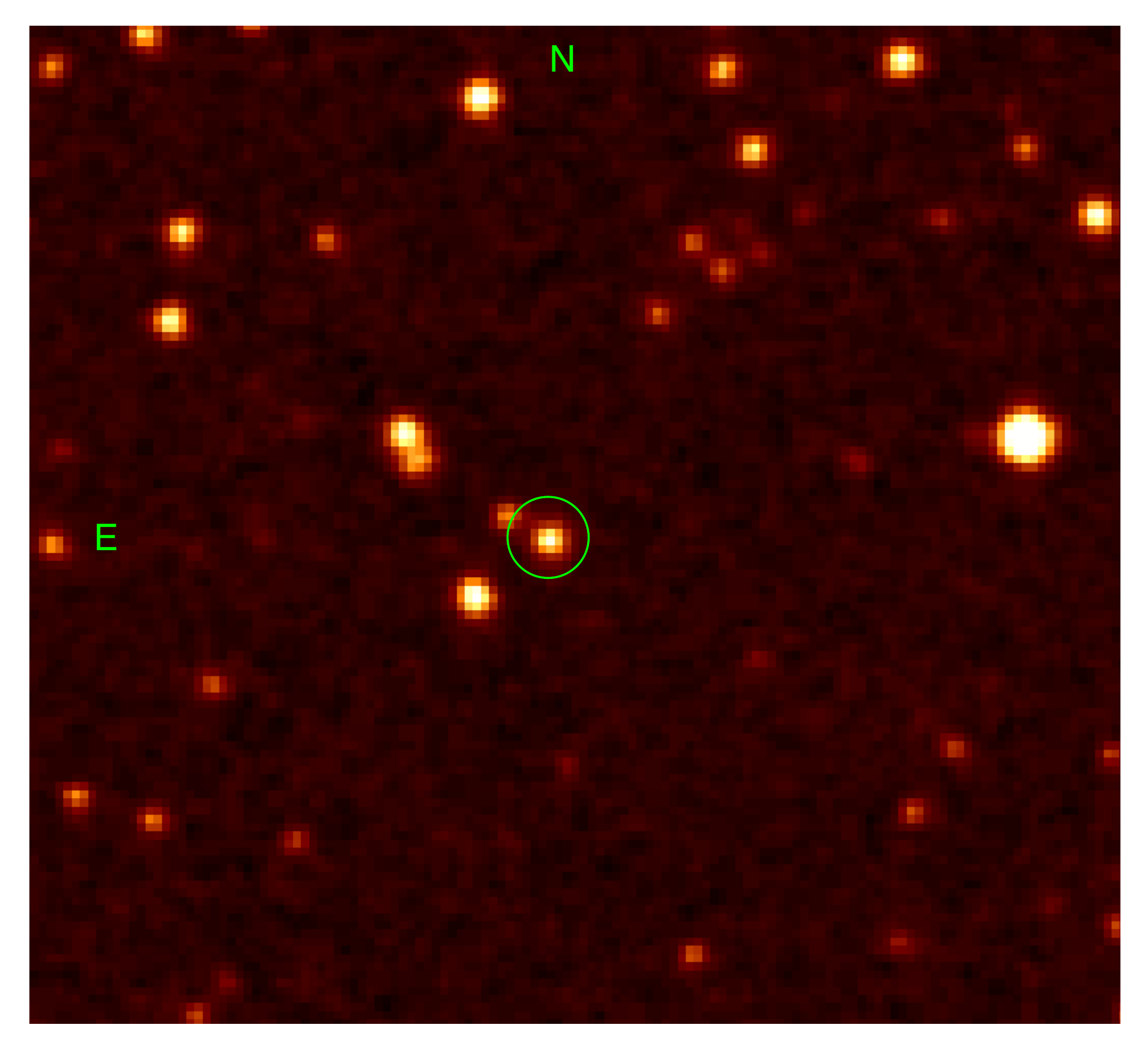}
\caption{DSS B-band image of region around PT Per. The green circle indicates the position of 3XMM J024251.2+564131. The size of the circle is much larger than the X-ray position uncertainty  (see text). The region shown is 2.5 arcmin. across.}
\end{figure}

The broad-band (0.2-12 keV) X-ray light curve obtained from the XMM-Newton observation is shown in Fig.3.  The light curve clearly shows four distinct minima occurring with a period of $\approx 4900$ seconds. Away from the minima the X-ray count rate shows variability by a factor $\sim 2$. 
\subsection{Optical observations}
After the discovery of the apparent periodic modulation in the X-ray light curve, time-resolved optical spectroscopy of PT Per was secured on the 4.2m William Herschel Telescope using the Intermediate dispersion Spectrograph and Imaging System (ISIS). Observations were made in twilight at high air-mass on three adjacent nights as detailed in Table 1. The observations used the R300B and R158R gratings, providing a spectral resolution of $\sim 3$ and $\sim 6$\AA\ for the blue and red arms, respectively, and giving useful wavelength coverage of $\lambda\lambda \approx 3800-9200$\AA. Table 2 summarises the spectrophotometric fluxes and equivalent broad-band magnitudes for these observations. The optical spectra of PT Per obtained with the WHT and the ISIS spectrograph were reduced using standard techniques in IRAF. The spectra were flux calibrated against standard stars observed immediately after the target exposures and wavelength-calibrated using arc spectra. The spectra were obtained in twilight with the target at high airmass ($>3$).
\begin{table}
\caption{Journal of WHT ISIS optical spectroscopic observations}
\begin{tabular}{llrr}
\hline
Observation & Date (MJD) & Exp.  & Slit  \\
& & time (s) & width (")\\
\hline

 CV1\_1& 57133.85639& 120 & 1 \\
 
 CV1\_2& 57133.85797&  120& 1  \\
 
 CV1\_3 &57133.85956&  120& 1  \\
 
 CV1\_4 &57133.86107&  120& 1\\
 
 CV1\_5 &57133.86259&  120& 1 \\
  
 CV1\_6 &57133.86409&  120& 1 \\
  
 CV1\_7 &57133.86561&  120& 1  \\
 
 CV1\_8 &57133.86711 & 120 & 1  \\
 
 CV1\_9 & 57133.86863 & 120 & 1 \\
 
 CV1\_10 & 57133.87013&  120& 1 \\
  
 CV2\_1 & 57134.85823 & 300& 2\\
  
 CV2\_2 & 57134.86175 & 300& 2 \\
  
 CV2\_3 & 57134.86534& 300& 2  \\
 
 CV2\_4 & 57134.86893 & 300& 2  \\
 
 CV2\_5 & 57134.87253 & 300& 2  \\
 
 CV3\_1 & 57135.85888 & 300& 2 \\
  
 CV3\_2 & 57135.86249 & 300& 2 \\
  
 CV3\_3 & 57135.86608 & 300& 2  \\
 
 CV3\_4 & 57135.86967 & 300& 2  \\
 
 CV3\_5 &57135.87327 & 300& 2  \\
 
\hline
\end{tabular}

\end{table}

\begin{table*}
\caption{Optical fluxes and magnitudes derived from the WHT ISIS observations}
\begin{tabular}{llrrrrrr}
\hline
Observation & Date (MJD)&$\log f_B$ &
  \multicolumn{1}{c}{B}&
  $\log f_V$ &
   \multicolumn{1}{c}{V} &
  $\log f_R$ &
   \multicolumn{1}{c}{R}\\
\hline
  Night 1 ave. & 57133 &-16.10 &  19.42$^{\mathrm{m}}$& -15.93&   19.14$^{\mathrm{m}}$&   -16.39 &  19.33$^{\mathrm{m}}$ \\
   Night 2 ave. & 57134 & -15.75 &  18.77$^{\mathrm{m}}$ &-15.67 &  18.26$^{\mathrm{m}}$ & -15.98  & 18.30$^{\mathrm{m}}$ \\
 Night 3 ave. & 57135 & -15.90   &18.88$^{\mathrm{m}}$ &-15.71  & 18.63$^{\mathrm{m}}$ & -16.13 &18.67$^{\mathrm{m}}$ \\
\hline
\multicolumn{7}{l}{Flux values are in units of erg cm$^{-2}$ s$^{-1}$}\\
  \end{tabular}

\end{table*}

\begin{figure}
\includegraphics[width=8.5cm]{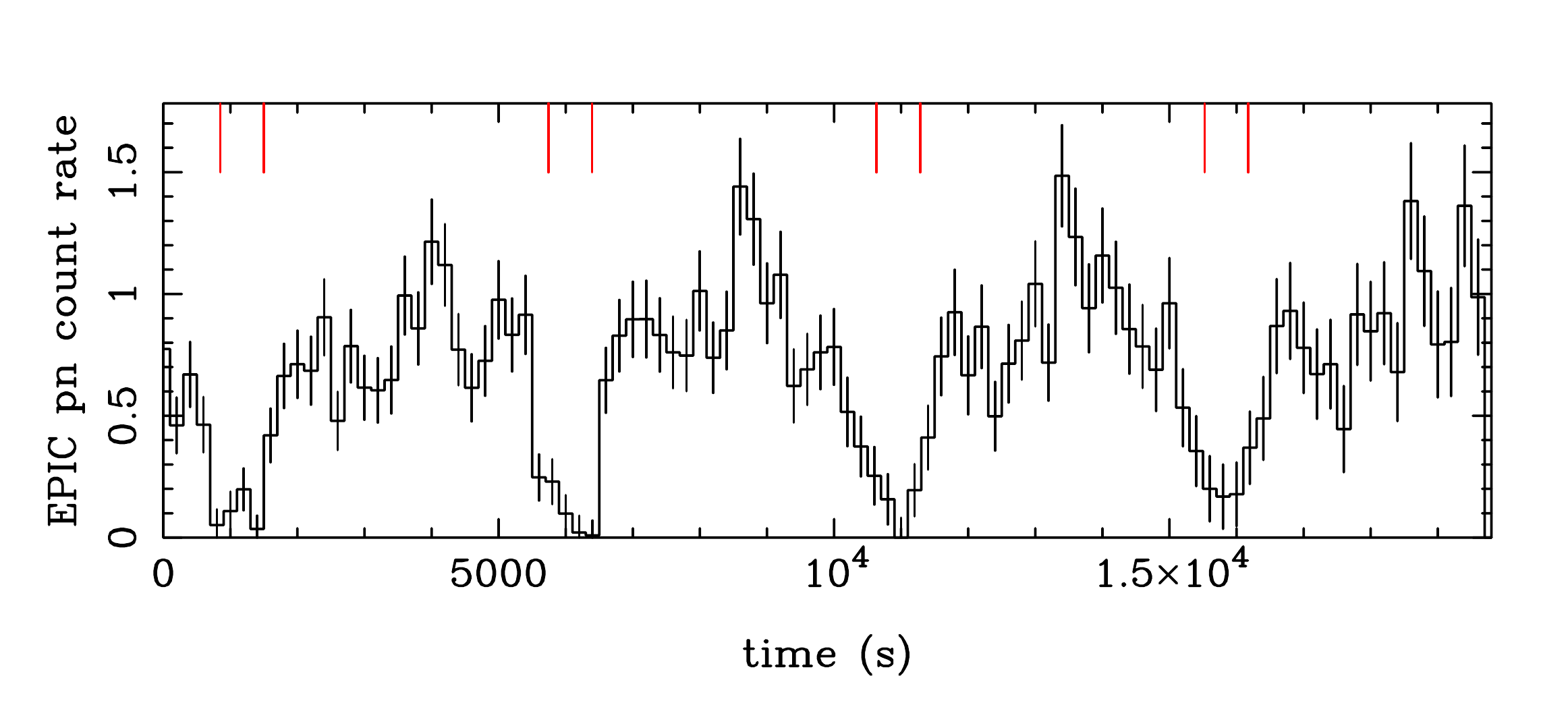}
\caption{The EPIC pn light curve of PT Per in the 0.2-12 keV band. The vertical red lines delineate the times of the minima in the light curve used to derive the minimum flux spectrum discussed in section 3.2. Time zero for this plot is MJD 55794.33363}
\end{figure}

\section{Analysis and results}
\subsection{X-ray timing}
The X-ray light curve of PT Per shown in Fig.3 shows four distinct minima which recur with a period of $\approx 4900$s. A Fourier transform of the XMM-Newton time series shows a highly significant peak at the corresponding frequency together with subsidiary peaks at the harmonics of this frequency, as expected given the shape of the light curve. On the basis of these data alone we estimate the modulation period to be $P=4900\pm 25$s (81.7 minutes, close to the CV minimum period). The error on the period is set by the time bin uncertainty in defining the occurrence of the first and last minima in the light curve. Fig.4 shows the X-ray light curve folded at $P=4900$s for the soft (0.5-2 keV) and hard bands (2-12 keV), together with the ratio of the hard to soft band intensities. The most obvious feature in the folded light curve is the broad minimum which has a phase width $\Delta\phi \approx 0.25$. The minimum is V-shaped in the soft band whilst the same feature in the hard band it is closer to being U-shaped.
{The ingress to this minimum appears to correspond to a possible softening of the count-rate ratio, accompanied by a possible hardening  in the centre of the minimum, although the significance of these features is low. }

The X-ray time series  of PT Per  was also extracted at much higher time resolution in order to search for additional periodicities in the data. No additional periods were found in the time series accumulated in 0.1s bins, ie. for periods $\ge$ 0.2s.

\begin{figure}
\includegraphics[width=8.5cm]{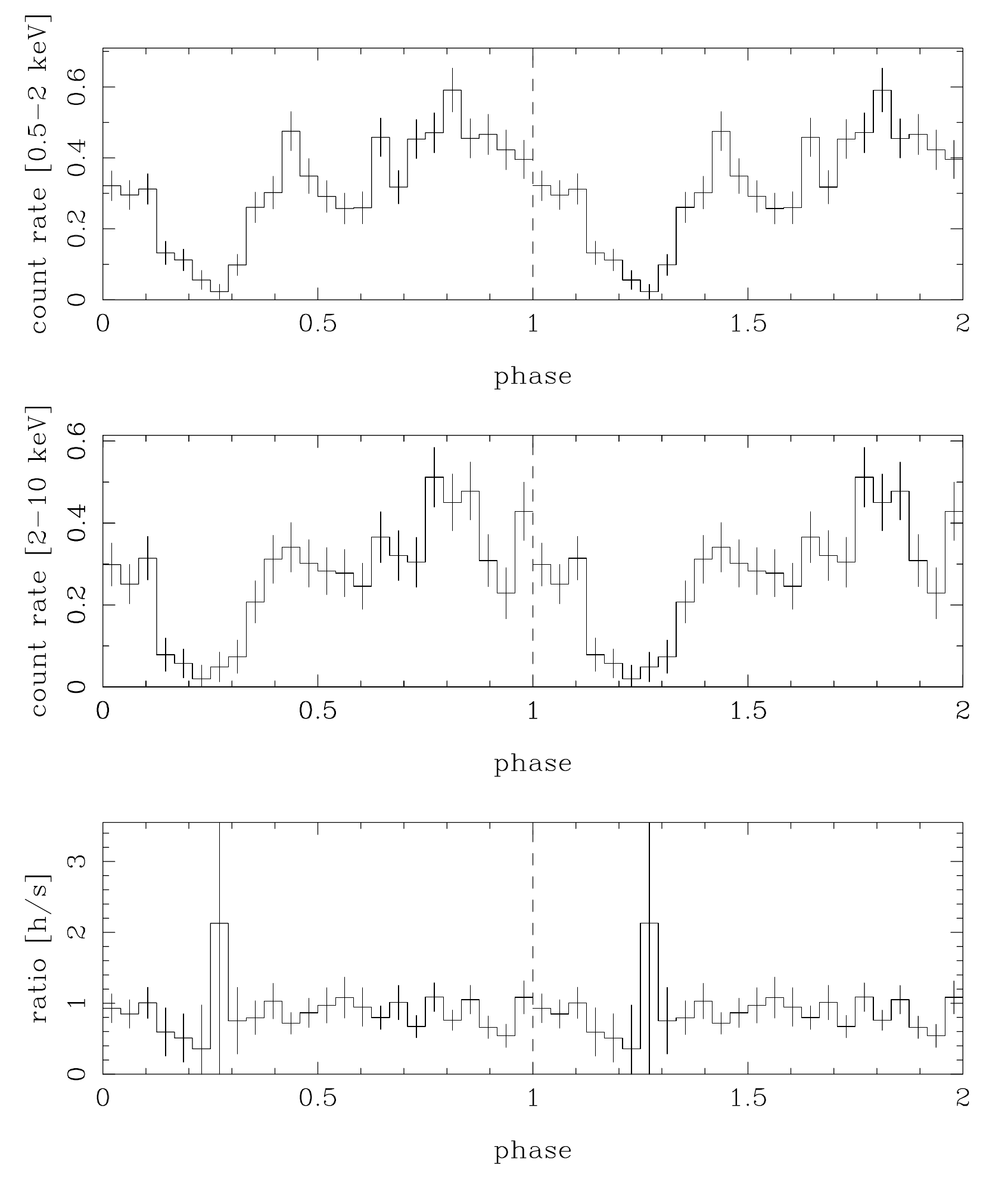}
\caption{X-ray timing data for PT Per folded at $P=4900$s. The top panel shows the soft band (0.5-2 keV) data, the centre panel the hard band (2-12 keV) data and the lower panel the ratio of hard band to soft band intensities. Phase zero for this plot is MJD 55794.33363}
\end{figure}
\subsection{X-ray spectra}
The X-ray spectrum of PT Per was extracted using standard methodology using SAS14.0\footnote{\url{http://xmm.esac.esa.int/sas/current/documentation/releasenotes/xmmsas\_14.0.0.shtml}} for the X-ray data reduction. Both EPIC pn and EPIC MOS2 spectra were extracted. Overall the spectrum is rather flat at low energies, suggestive of low absorption as expected for an object in the Galactic anti-centre, and a relatively hard featureless continuum except for a prominent emission line at $\approx 6.7$ keV.
The derived source and background spectra were fitted using XSPEC12.0\footnote{\url{http://heasarc.gsfc.nasa.gov/docs/xanadu/xspec/manual/
XspecIntroduction.html}} and a variety of spectral models: a simple power law, an optically thin thermal and a model combining power law and thermal components, and lastly a partially covered thermal model. In each case the spectra were modelled including line-of-sight photoelectric absorption. The results of the spectral fitting are shown in Table 2. Both power law and thermal spectral models give acceptable fits, but they fail to fit the data around 6.7 keV (this line is centred at $E=6.65\pm 0.04$keV and is presumably Fe K$\alpha$). A better fit is achieved either with the model with both power law and thermal components, or with a partially covered thermal model as shown in Fig.5. These latter models also give an adequate fit to the Fe line, due to the lower temperature of the thermal component. The F-test shows that these latter two models are a statistically better representation of the data. 

The X-ray spectrum of PT Per was also extracted for the time periods corresponding to the minima in the light curve (marked with red lines in Fig.3). The resultant spectrum is very noisy due to the low number of counts. The best fit was achieved for a thermal spectrum with $kT\approx 1.3$ keV, in accord with the possible softening noted in the folded light curve (Fig.4). There is no significant change in the fitted column density value.
\begin{table*}
{\centering \caption{Spectral fitting results}}
\begin{tabular}{lcccrl}
\hline
Model & $N_H$& $\Gamma$ & $kT$ & red.$\chi^2/\mathrm{dof}$& Note\\
           &$\times 10^{20}\mathrm{cm^{-2}}$ &Photon index  & keV & &\\
\hline
Power law & 6.6$\pm$0.9&1.31$\pm$0.05 &- &1.24/166\\
Thermal & 5.9$\pm$0.7&- &34.3$\pm$16 &1.25/166\\
Power law + thermal & 6.3$\pm$0.8 & -0.07$\pm$0.5& 6.52$\pm$1.5 &1.13/164 \\
Partially covered thermal & 6.5$\pm 0.7$& -&6.4$\pm0.9$& 0.93/160& Partial covering $N_H=5.7 \pm 1.6\times 10^{22}\ \rm cm^{-2}$\\
& & & & &  with covering fraction=$0.5\pm 0.06 $\\
\hline
\end{tabular}

\end{table*}
\begin{figure}
\includegraphics[width=8.5cm]{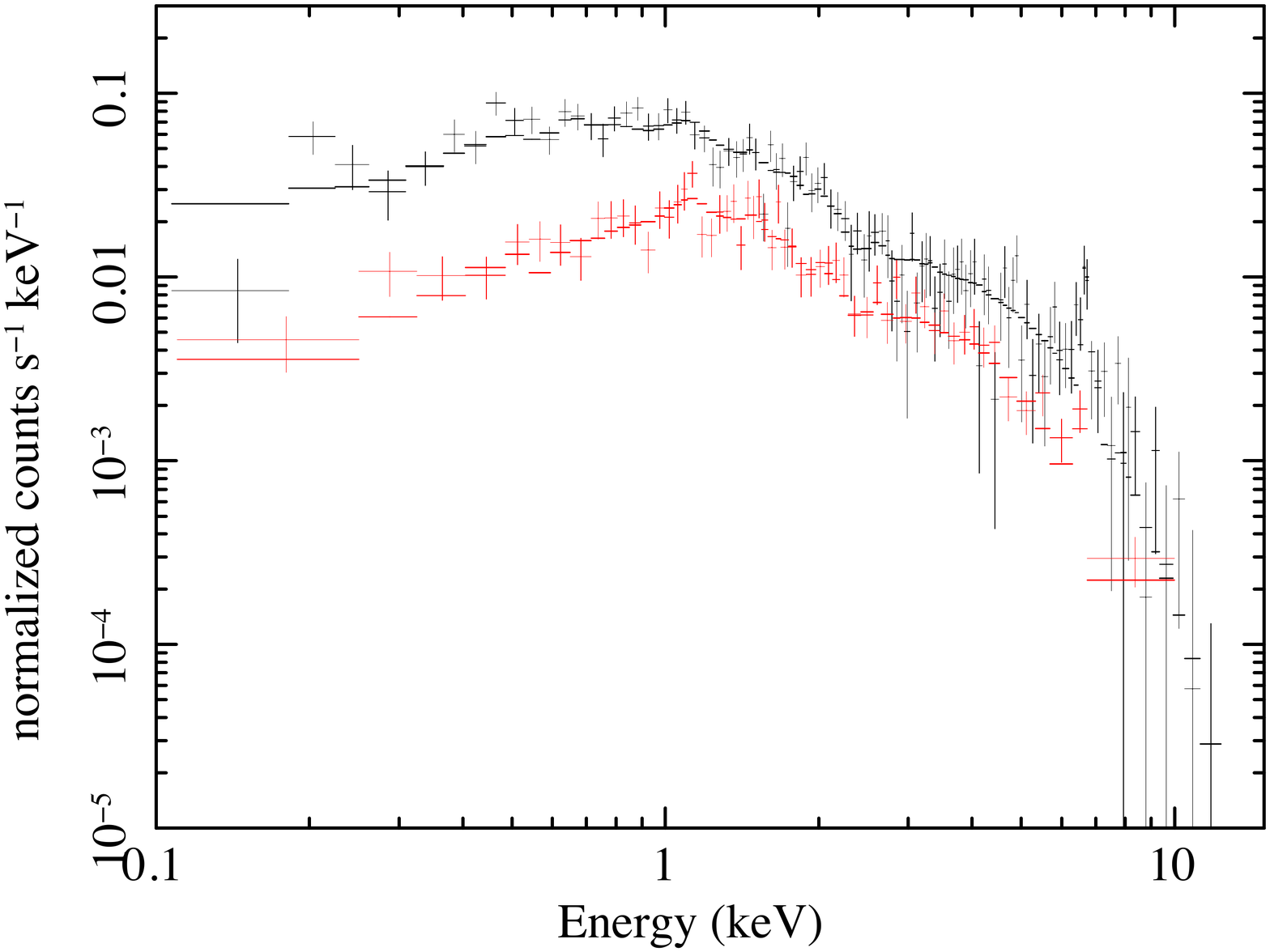}
\caption{Spectral fit to the EPIC pn (black) and MOS2 (red) data with a partial covering thermal model, together with low energy photoelectric absorption.}
\end{figure}
\subsection{Optical spectra}
The time-averaged spectra taken on each separate night (see Table 1) are shown in Fig.6.  Due to the high airmass of the observations, there is some systematic uncertainty in the absolute flux values as varying slit losses may have occurred between the target and standard star observations (see section 3.4). There is also a small difference in the slit losses (within each observation) between the blue and red arms of the spectrograph producing a change in flux normalisation around $\lambda=5400$\AA, as can be seen most clearly in the spectra obtained on the first two nights. The observing conditions also explain the relatively high noise in the spectra. The observed fluxes in the spectra correspond to B$\approx 19^{\mathrm{m}}$,  V$\approx 18.6^{\mathrm{m}}$ and R$\approx 18.6^{\mathrm{m}}$.

Overall the spectra show a relatively featureless blue continuum with no prominent emission or absorption features. There appears to be no sign of the secondary star in the optical spectrum, assuming that this is a late-type dwarf. Some apparent flux variability by a factor 2-3 between nights is evident (but see section 3.4). In Fig.6 the positions of the Balmer lines are marked. There is some evidence of the Balmer lines in emission in the spectra. A narrow  H$\alpha$ line is seen in the spectra for nights 1 and 3. Broader features are possibly present at H$\beta$ and H$\gamma$. There is also evidence of broad and blue-shifted Balmer absorption in the spectra, but again this does not appear to be constant. An additional emission line is present in two of the spectra at $\lambda \approx 6300$\AA\ is presumably the [OI] sky line.  On shorter timescales the individual spectra obtained also show evidence for variability, with the lines detected in the time-averaged spectra varying considerably from exposure to exposure, but the equivalent widths even for H$\alpha$ are only a few \AA\ (comparable with the errors in many cases), precluding a detailed analysis. We also note that there appear to be wavelength shifts for the narrow H$\alpha$ line by as much as $\pm 6$\AA, equivalent to $\sim \pm300\ \mathrm{km\ s^{-1}}$.


\begin{figure}
\includegraphics[width=8.5cm]{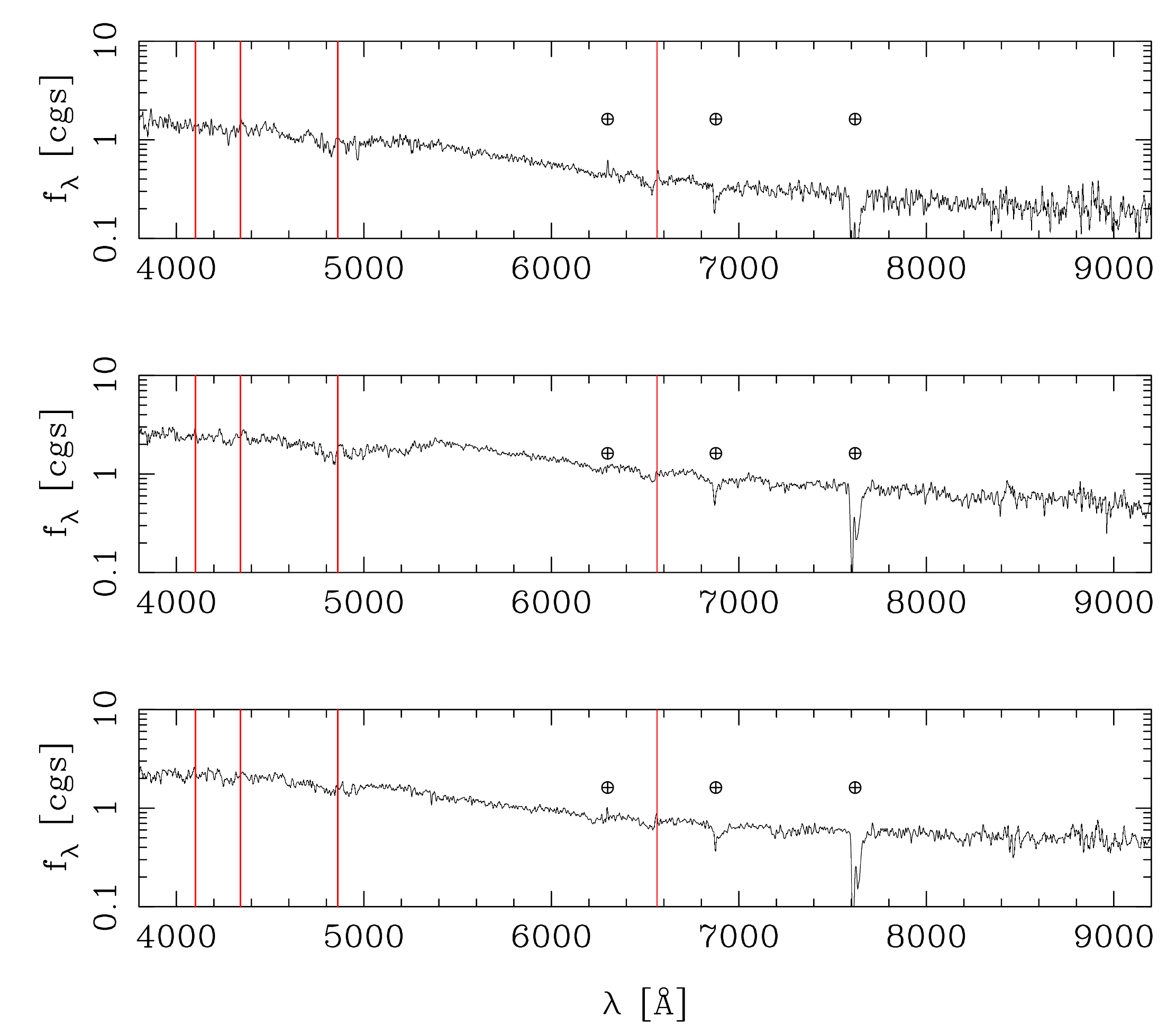}
\caption{Time-averaged optical spectrum of PT Per for the observations taken on three nights (see Table 1; first night at the top of figure). Telluric features are marked. The red lines mark the wavelengths expected for H$\delta$, H$\gamma$, H$\beta$ and H$\alpha$ respectively from left to right. Flux values are in units of $10^{-16}\ \mathrm{erg\ cm^{-2}\ s^{-1}\ \AA^{-1}}$.}
\end{figure}
\subsection{Optical spectrophotometry}

{The time-resolved spectra listed in Table 1 were used to derive fluxes in the standard BVR bands and then folded on the $P=4900$s period. Note that the phase cannot be connected to that of the X-ray modulation. Examination of the light curves shows that the flux appears to change by a factor $\sim 3$ with 
 only detailed differences between bands. Because the observations were made at high airmass it is not possible to be confident of the flux calibration between nights and it is likely that differential slit losses or transparency variations may be contributing to the apparent variability. The derived light curves fortuitously cover almost a full cycle of the $P=4900$s period (on different days of course), but show flux discontinuities which are unlikely to be real. Removing the discontinuities (which requires a flux renormalisation between nights 1 \& 3 by factors $\approx$2, 1.3 \& 1.1 in the B, V \& R bands respectively) nevertheless leaves a significant modulation of the optical flux with time or phase.}
\section{Discussion}
\subsection{Nature of the periodic modulation}
The results presented in section 3 broadly support the idea that PT Per is a cataclysmic variable (CV). Its X-ray light curve appears to be strongly modulated with a period of 4900s (82 minutes) and resembles that of magnetic CVs at X-ray wavelengths, eg. \citet{warner03}. The X-ray light curve minimum has a phase width $\Delta\phi\approx 0.25$. Its X-ray spectrum is relatively hard, shows a 6.7 keV Fe line, and may display variability with phase of the 82-minute period in as far as the light curve minimum is softer (at low significance). Its optical spectrum however is less easy to interpret. The optical continuum shape is similar to spectra of dwarf novae in their high states \citep{warner03}, where the accretion disc dominates the light and the Balmer lines are very weak, leading to a $f_{\mathrm{opt}}\propto \lambda^{-2}$ \citep[eg.][]{fkr}.

The most obvious interpretation which fits most of the observational data is that the period is orbital in nature with the minima being associated with either an occultation of the white dwarf (WD) by the secondary star in a high inclination system, or associated with the disappearance of the X-ray emission region (ie. the accretion column) on a highly magnetised WD  which is co-rotating with the orbital period, i.e. a polar system. The first possibility can be ruled out for any reasonable system parameters as the eclipse is much too long for an sensible choice of secondary star radius \citep[cf. eclipses in eclipsing dwarf novae like HT Cas have 
$\delta\phi<0.1$, eg.][]{nucita09}. The second possibility thus seems more likely to be the correct explanation and the other properties of the system such as the  X-ray spectral fit and parameters are also features which may be expected in a polar system by analogy with other objects in this class, eg. \cite{warner03}. We note however that polars often have an ultra-soft component in the spectrum \citep[although such a component is not evident in around 30\% of systems, eg. ][]{ramsay04}; there is no evidence for this in PT Per. The spectral softening seen during the light curve minima is crudely consistent with our ideas of the structure of the accretion column.

What is completely unexpected for a polar however is the optical spectrum. Essentially all magnetic CV systems (and indeed most dwarf novae) have optical spectra which show very strong H emission lines in their spectra. In contrast the optical spectra of PT Per have very weak H Balmer lines and instead show a mostly featureless blue continuum. This mismatch with expectations could of course be due to time variability, as the X-ray observations were made in 2011 and the optical observations were made in 2015. One could thus hypothesise that PT Per has changed from a polar to a non-magnetic CV in the space of 4 years. under the assumption that the 2015 optical spectra are dominated by an accretion disc. This is, however, difficult to understand as the polar nature of a system is very hard to modify on short timescales. In polar systems the WD magnetic field is strong enough to impose phase-locking between the WD and the rotation of the binary and the field controls the accretion flow from the secondary so that the system has no accretion disc. 

{ The 2015 optical spectra could however be dominated by the WD in the system. Although the distinctive broad absorption lines expected are not immediately apparent, there are some shallower features which might represent Zeeman components.
In order to investigate this possibility we have computed the H$\alpha$, H$\beta$ and H$\gamma$ Zeeman components for a range of assumed WD magnetic fields. In each case the Zeeman components for a given field strength are weighted according to size of visible area for an assumed inclination angle of $i=15-20^\circ$. A centred dipole field is assumed resulting in a factor 2 variation of field strengths from the pole
to the equator and (implicitly) a homogeneous temperature distribution. 
In Fig.7 we show the wavelength shifts as a function of assumed equatorial field for $B_{eq}$ in the range 10-30 MGauss for H$\alpha$, H$\beta$ and H$\gamma$ Zeeman components. As can be seen the complex set of shifts are a fast function of the magnetic field, implying that identifying one or more of the discrete components can potentially tie down the white dwarf field. 

In Fig.8 we show the match between the prediction and the observed spectrum for an assumed field of $B_{eq}=15$ Mgauss, equivalent to a polar field $B_p\approx25-27$  MGauss for an as the assumed ratio between $B_{p}$ and $B_{eq}$ will be 1.6-1.7 for our assumed geometry. Overall the prediction seems to give a reasonably good representation of the data, in particular there is a good match for the H$\alpha$ component at $\lambda \approx 6530$\AA, the H$\beta$ components between $\lambda 4700-5000$\AA\ and the H$\gamma$ feature at $\lambda 4300$\AA. Some features such as the absorption component of H$\alpha$ at $\lambda \approx 6000 - 6300$\AA\ do not fit well. The Zeeman components will of course also be smeared by orientation effects as the WD rotates \citep[eg.][]{wick00} and the assumption of a homogeneous WD is unlikely to be valid, so it would be surprising to get a perfect match.
 This analysis suggests that PT Per has a polar field of the order $B_p\approx 25-27$ MG. This is a typical value for a magnetic CV \citep[eg.][]{wick00}.
This exercise is intended as an illustration of where we expect absorption in a dipolar Zeeman atmosphere. It is not possible to comment on the depth of the Zeeman features in the spectrum as such an analysis requires a radiative transfer computation that is not warranted by the current data quality. 

We can only give crude constraints on the WD temperature. Comparison of the continua of the spectrum of PT Per with the single WD SDSS J101805.04+011123.52 \citep{kep13} suggests $T_{\mathrm{eff}}\sim 10,000$ K. A stronger constraint on the temperature requires broader wavelength coverage, but unfortunately PT Per is not covered in the GALEX survey or by the SWIFT UVOT.\footnote{\url{http://galex.stsci.edu/GR6/};  \url{http://www.swift.ac.uk}}

The analysis of the 2015 spectra of PT Per thus suggests that we are seeing PT Per in a low state where accretion is at a minimum and optical light from the accretion columns and accretion flow is essentially absent. The narrow H$\alpha$ line could originate at the secondary star in which case the wavelength shifts could be due to the orbital motion. For reasonable system parameters, e.g $M_1=0.7M\sun$, $M_2\approx 0.075M\sun$ (see section 4.2), the expected radial velocity amplitude for the secondary is  $v\sim200\ \mathrm{km\ s^{-1}}$, corresponding to $\Delta\lambda \sim 6$ \AA\ for H$\alpha$ (for an assumed inclination $i=20^\circ$).}

The lack of an apparent accretion stream eclipse and eclipse by the secondary  also provide constraints on the system geometry. The bright phase duration suggests $\tan(i) = 1.24/\tan(\delta)$, ie. $\tan(i) = -1/(\cos(2\pi \Delta\phi_B)\tan(\delta))$, where $i$ is the inclination, $\delta$ is the co-latitude of the magnetic pole and $\Delta\phi_B$ is the fraction of the cycle which is bright. The lack of eclipse due to the stream implies $i < \delta$; with the previous constraint this means $i < 49^\circ$.

{To put these findings in context, we note that polars probably spend half of their time in quiescence \citep[eg.][]{wu08,kal12}, so finding PT Per in a high state in 2011 and in quiescence in 2015 is not surprising. What is notable however is the complete lack of any signs of accretion onto the WD in the 2015 spectra. Most studies of polar low states to date, eg. \citet{beu00}, \cite{mas00}, but also see \cite{sch93a} and \cite{beu07}, show optical spectra that still contain prominent H emission lines (and often cyclotron features as well); these are absent in our spectra of PT Per.}

\begin{figure}
\includegraphics[width=8.5cm]{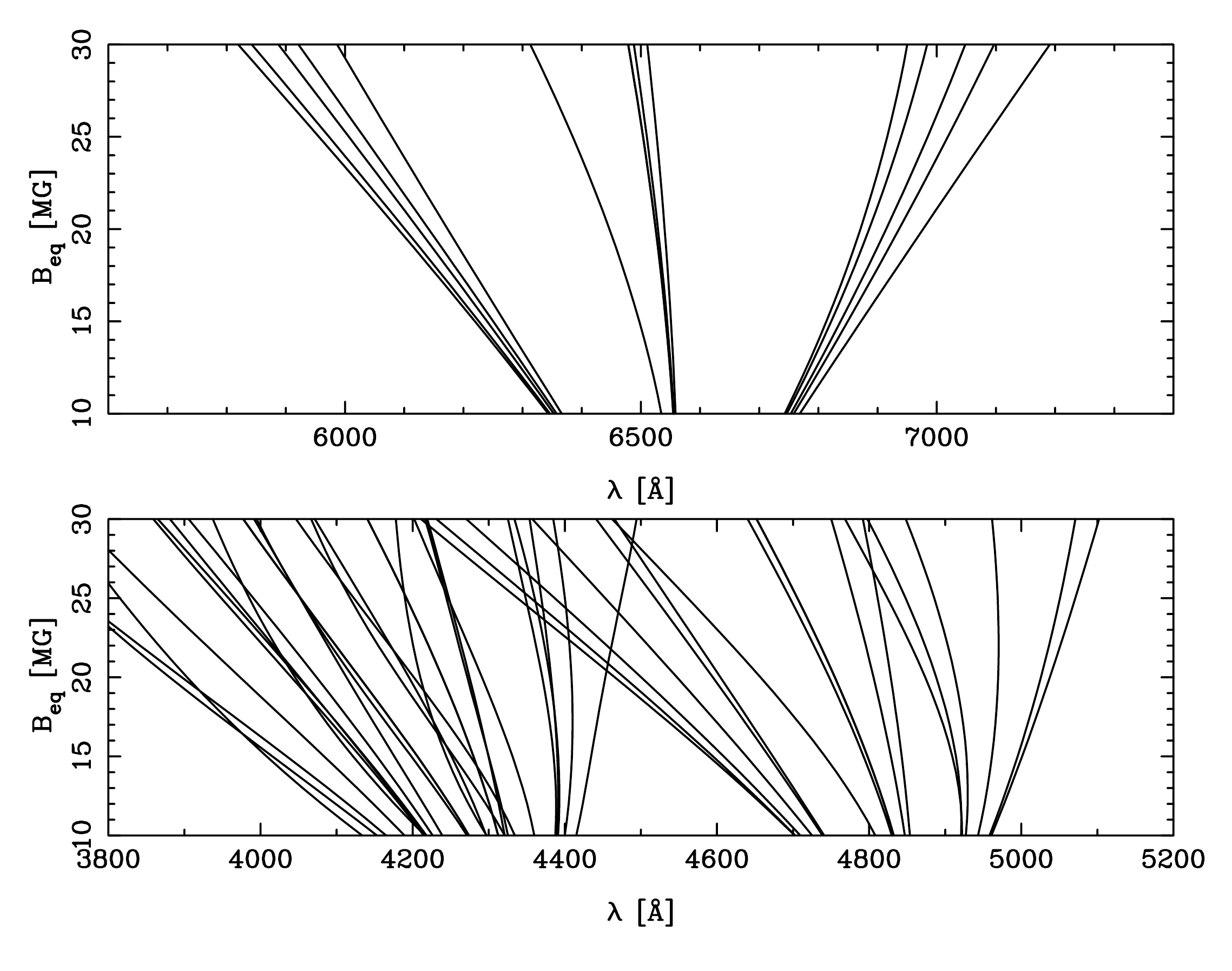}
\caption{Zeeman line shifts for H$\alpha$, H$\beta$ and H$\gamma$ as a function of assumed equatorial field strength.}
\end{figure}

\begin{figure}
\includegraphics[width=8.5cm]{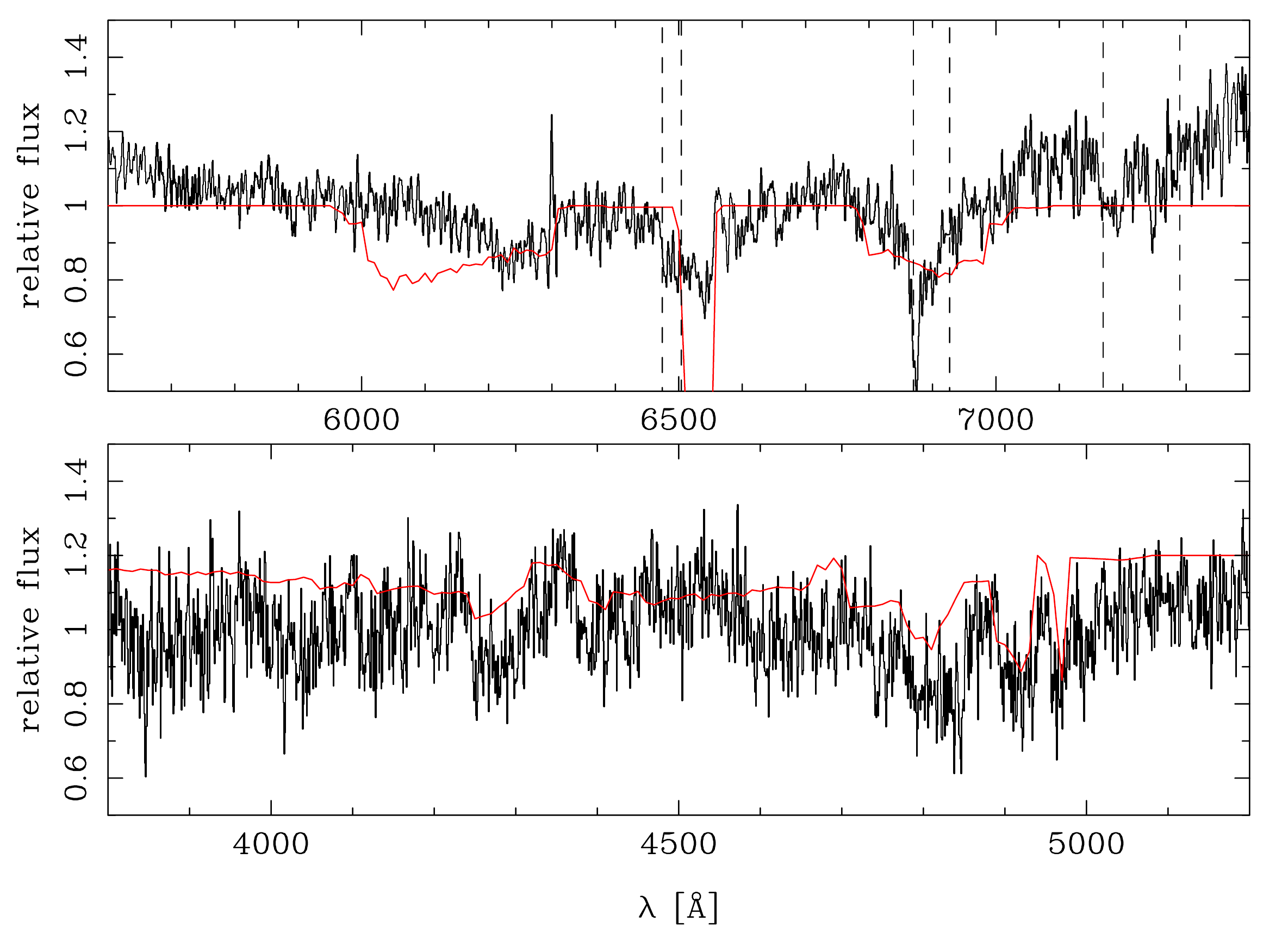}
\caption{ Comparison of the average optical spectrum of PT Per for the observations taken on all three nights with the predicted Zeeman absorption spectrum for a polar field of 25 MG (for details, see text). The portions of the optical spectrum shown have been normalised by a linear fit to the continuum. The dashed lines indicate the parts of the spectrum potentially affected by incompletely removed atmospheric absorption.}
\end{figure}

\subsection{System parameters}
Assuming the 2015 optical spectrum is dominated by a high-field magnetic WD in the PT Per system, we can derive a photometric distance from its apparent magnitude and colours. The WD has $\rm V\approx 18.6^{\mathrm{m}}$ and B-V$\approx 0.4^{\mathrm{m}}$. From the compilation of WD system parameters of \cite{mccook99}, the typical  absolute magnitude for B-V$\approx 0.4^{\mathrm{m}}$ is 
$\rm M_V\approx 13.7^{\mathrm{m}}$, giving a distance modulus$\approx 4.9^{\mathrm{m}}$ and a distance $d\approx 90$ pc. 
Alternatively, we can compare the WD in PT Per with WD1544-377 in \cite{holberg08}. This WD has $T_{\mathrm{eff}}\approx 10500$ K and is at distance of 15 pc, implying $M_{\rm V}=13.66^{\mathrm{m}}$, essentially identical with the value adopted above. As PT Per is at Galactic coordinates $ l_{\mathrm{II}}=137.8^{\circ}, b_{\mathrm{II}}=-2.92^\circ$, the distance estimate gives a height above the Galactic plane of $\approx 5$ pc. For $d\approx 90$ pc, the X-ray luminosity of PT Per is $L_{\mathrm X}\approx 3.4\times 10^{30}\ {\mathrm{ erg\ s^{-1}}}$. This is a relatively low luminosity for a magnetic CV system, in fact closer to what is typically observed in dwarf novae and nova-like variables, and may imply a low mass-transfer rate in this system. {Magnetic WDs have higher masses and hence smaller radii than single WDs \citep{ferr15}; this would imply that our distance and luminosity estimates may be too low.}

{ If we assume the modulation period of 82 minutes is the orbital period, the system is expected to have the secondary star filling its Roche lobe and we can directly estimate the mass and radius of the secondary star from the empirical donor sequence of \citet{kni06}, \citep[see also][]{kni07} which gives spectral type L3.5, $M_2=0.074 {\mathrm{M}_{\sun}}$ and an absolute magnitude $M_{\rm V}=21.25^{\mathrm{m}}$. With the distance estimate derived above, this predicts $V\approx 26^{\mathrm{m}}$, implying that a negligible amount of the V-band flux would be due to the secondary star, consistent with the optical spectrum which does not show any unambiguous sign secondary at the longest wavelengths (see Fig.6). The 2MASS near-IR colours for PT Per, J-H=0.5 and H-K=0.16 \citep{cut03}, are broadly consistent with other polars \citep[eg.][]{hoa02} although rather bluer in H-K and do not provide any additional constraints on the secondary star, especially as cyclotron emission may make a contribution in these bands.}

With an orbital period of 82 minutes, PT Per is a candidate ``period bounce" system. Period-bounce systems are objects below the minimum mass for hydrogen burning where the mass transfer timescale is shorter than the thermal timescale, leading to a period {\bf increase} as the system loses mass, ie. in which the secondary has become a brown dwarf with degenerate core, eg. \cite{kolb99}. This would explain the underluminous nature of the secondary (as core nuclear burning would have ceased).

%


\section{Conclusions}
We have presented a study of PT Per based on archival XMM-Newton X-ray data and new optical spectroscopy from the WHT with ISIS. 
From the observational data  it seems most likely that PT Per is a polar system with the X-ray light curve modulation being due to the periodic disappearance of the accretion column on a magnetised WD. The depth and phase width of the light curve minima provide constraints on a combination of the emission region size, its magnetic co-latitude and the orbital inclination of the system. Modelling the light curve to determine these parameters would be warranted when higher quality data become available. The absence of strong emission lines in the 2015 optical spectra can be due to a low accretion state where the bright emission lines produced by X-ray photoionisation of the gas flow onto the WD would be faint and the spectrum becomes dominated by the WD,  although it is true to say that the line features in PT Per are very weak even when compared with other polar systems in their low states; examples include \cite{sch93b} \& \cite{sch95} where, even in the absence of emission related to accretion, H$\alpha$, likely due to coronal activity on the donor star, is present. }The spectra are broadly consistent with a low temperature and high field WD. The absence of any secondary star features in the optical spectra require the secondary to be very underluminous, consistent with PT Per being a period-bounce system in which the secondary has become a brown dwarf with a degenerate core. This would also explain the low observed X-ray luminosity as period-bounce systems have low accretion rates. Such systems, although believed to be common, are difficult to discover observationally due to their low accretion rates and thus luminosities.

Future studies of PT Per are required to confirm the polar nature of the system. In particular optical polarimetry will provide unambiguous evidence of the magnetic nature of the white dwarf, whilst time-resolved spectroscopy, optical photometry and more extensive X-ray coverage will allow the orbital ephemeris to be determined, constrain the accretion geometry and WD magnetic field and its effective temperature. Given the limited quality of present data, it is also possible that PT Per is an intermediate polar rather than a polar. This can be resolved by optical spectroscopy or photometry which would reveal two modulation periods if it is an intermediate polar.

\section*{Acknowledgements}
Based on observations obtained with XMM-Newton, an ESA science mission 
with instruments and contributions directly funded by 
ESA Member States and NASA. Based on observations made with the WHT operated on the island of La Palma by the Isaac Newton Group in the Spanish Observatorio del Roque de los Muchachos of the Instituto de Astrofísica de Canarias. {We acknowledge with thanks the variable star observations from the AAVSO International Database contributed by observers worldwide and used in this research.}

\bibliography{ptperpaper}
\label{lastpage}
\end{document}